\documentclass[pra,onecolumn,superscriptaddress]{revtex4-1}

\usepackage{amsmath}
\usepackage{braket}
\usepackage{mathtools}

\usepackage{pgfplots}

\usepackage{hyperref}
\usepackage{lipsum}
\usepackage{color}
\usepackage{comment}
\usepackage{bm,amssymb,graphicx,braket,dsfont,mathtools,bbold,MnSymbol}
\DeclareMathOperator{\Tr}{Tr}

\usepackage[latin1]{inputenc}
\usepackage[caption=false]{subfig}
\usepackage{pgfplots}
\usepackage{cancel}
\usepackage{tabularx,booktabs}
\usepackage{multirow}
\usepackage{fancyhdr}
\usepackage{amsthm}
\usepackage{inputenc}

\newcommand{\be}{\begin{equation}}
\newcommand{\ee}{\end{equation}}
\newcommand{\ba}{\begin{align}}
\newcommand{\ea}{\end{align}}
\newcommand{\sysb}{\left\{\begin{array}}
\newcommand{\syse}{\end{array}\right.}
\newcommand{\baa}{\begin{array}}
\newcommand{\eaa}{\end{array}}
\newcommand{\bs}{\begin{split}}
\newcommand{\es}{\end{split}}

\newcommand{\matb}{\left(\begin{array}}
\newcommand{\mate}{\end{array}\right)}

\newcommand{\lt}{\left(}
\newcommand{\rt}{\right)}

\newcommand{\nol}{\nonumber \\}

\begin{document}

\title{Open quantum generalisation of Hopfield neural networks}

\author{P. Rotondo}
\affiliation{School of Physics and Astronomy, University of Nottingham, Nottingham, NG7 2RD, UK}
\affiliation{Centre for the Mathematics and Theoretical Physics of Quantum Non-equilibrium Systems,
University of Nottingham, Nottingham NG7 2RD, UK}
\author{M. Marcuzzi}
\affiliation{School of Physics and Astronomy, University of Nottingham, Nottingham, NG7 2RD, UK}
\affiliation{Centre for the Mathematics and Theoretical Physics of Quantum Non-equilibrium Systems,
University of Nottingham, Nottingham NG7 2RD, UK}
\author{J. P. Garrahan}
\affiliation{School of Physics and Astronomy, University of Nottingham, Nottingham, NG7 2RD, UK}
\affiliation{Centre for the Mathematics and Theoretical Physics of Quantum Non-equilibrium Systems,
University of Nottingham, Nottingham NG7 2RD, UK}
\author{I. Lesanovsky}
\affiliation{School of Physics and Astronomy, University of Nottingham, Nottingham, NG7 2RD, UK}
\affiliation{Centre for the Mathematics and Theoretical Physics of Quantum Non-equilibrium Systems,
University of Nottingham, Nottingham NG7 2RD, UK}
\author{M. M\"uller}
\affiliation{Department of Physics, Swansea University, Singleton Park, Swansea SA2 8PP, UK}

\begin{abstract}
We propose a new framework to understand how quantum effects may impact on the dynamics of neural networks. We implement the dynamics of neural networks in terms of Markovian open quantum systems, which allows us to treat thermal and quantum coherent effects on the same footing.  In particular, we propose an open quantum generalisation of the Hopfield neural network, the simplest toy model of associative memory. We determine its phase diagram and show that quantum fluctuations give rise to a qualitatively new non-equilibrium phase. This novel phase is characterised by limit cycles corresponding to high-dimensional stationary manifolds that may be regarded as a generalisation of storage patterns to the quantum domain.
\end{abstract}

\maketitle

\section{Introduction}

Neural networks (NNs) \cite{Haykin:book} - artificial systems inspired by the neural structure of the brain - have become essential tools for solving tasks where more traditional rule-based algorithms fail.  Examples are pattern and speech recognition \cite{Bishop:ML:2006}, artificial intelligence \cite{Mnih:Nat:2015, Silver:Nat:2016}, and the analysis of big data \cite{Hinton:Science:2006}. All these NNs evolve according to the laws of classical physics.

It is widely believed that computational processes can benefit by exploiting the properties of quantum mechanics. %such as coherence and entanglement.  
Seminal works by Shor \cite{Shor:SIAM:1999} and Grover \cite{Grover:PRL:1997} have proved the existence of quantum algorithms that systematically outperform their classical counterparts. More recently, there has been a growing interest in proposing and realizing quantum architectures that may be possibly used in the future as quantum computers: for instance, Shor's algorithm for the factorization of integer numbers has been implemented on a small 11-qubits trapped-ion system \cite{Monz:Science:2016} and D-Wave machines, based on superconducting qubits, have shown potential speedup in solving a particular class of NP-hard problems via quantum annealing \cite{Denchev:PRX:2016}. Although the accurate control of large qubit registers (the scalability challenge) is non-trivial and the computational speedup of the latter is still debated \cite{mandra:arXiv:2016, Nishimori:arXiv:2016,PhysRevLett.119.110502}, these represent a concrete step towards large-scale quantum information processors.

An important and timely question is whether it is possible to take advantage of quantum effects in NN computing \cite{Biamonte2016quantum}. To our knowledge, only a few studies exist on the learning capacity of quantum perceptron models \cite{Lewenstein:JMO:1994,Wiebe:arXiv:2016} and on quantum Boltzmann machines \cite{Amin:2016:arXiv}. To date, however, none of the existing proposals \cite{ventura2000quantum, ezhov2000quantum, gupta2001quantum} allows to define a satisfactory framework to consider quantum effects in NN dynamics. The problem is a conceptual one: the dynamics of closed quantum systems is governed by deterministic temporal evolution equations, whereas NNs are always described by dissipative dynamical equations, thus preventing any straightforward generalization of NNs computing in quantum systems \cite{Schuld:QInf:2014}. Here we overcome this obstacle, by proposing a framework for quantum NNs based on open quantum systems (OQSs). We consider, in particular, the conceptually simplest case of Markovian dynamics, where the evolution of the density matrix is described by a Lindblad equation \cite{Breuer:book}.
%In most cases correlations in the dissipative environment are sufficiently short-lived to allow a description in terms of a Markovian master equation in Lindblad form \cite{Breuer:book}.
%What we aim to show is that the open nature of quantum systems can be exploited as a resource for OQS based NNs, by careful combination of coherent and dissipative effects. The framework for open quantum NNs we introduce here is close in spirit to that of recent works where environment engineering is exploited to recover the result of a quantum computation from the dark state of a dissipative dynamics \cite{Diehl:NatPhys:2009, Verstraete:NatPhys:2009, Caspar:PRA:2016, Schindler:NatPhys:2013}.

\begin{figure}[t]
\includegraphics[width=0.45 \textwidth]{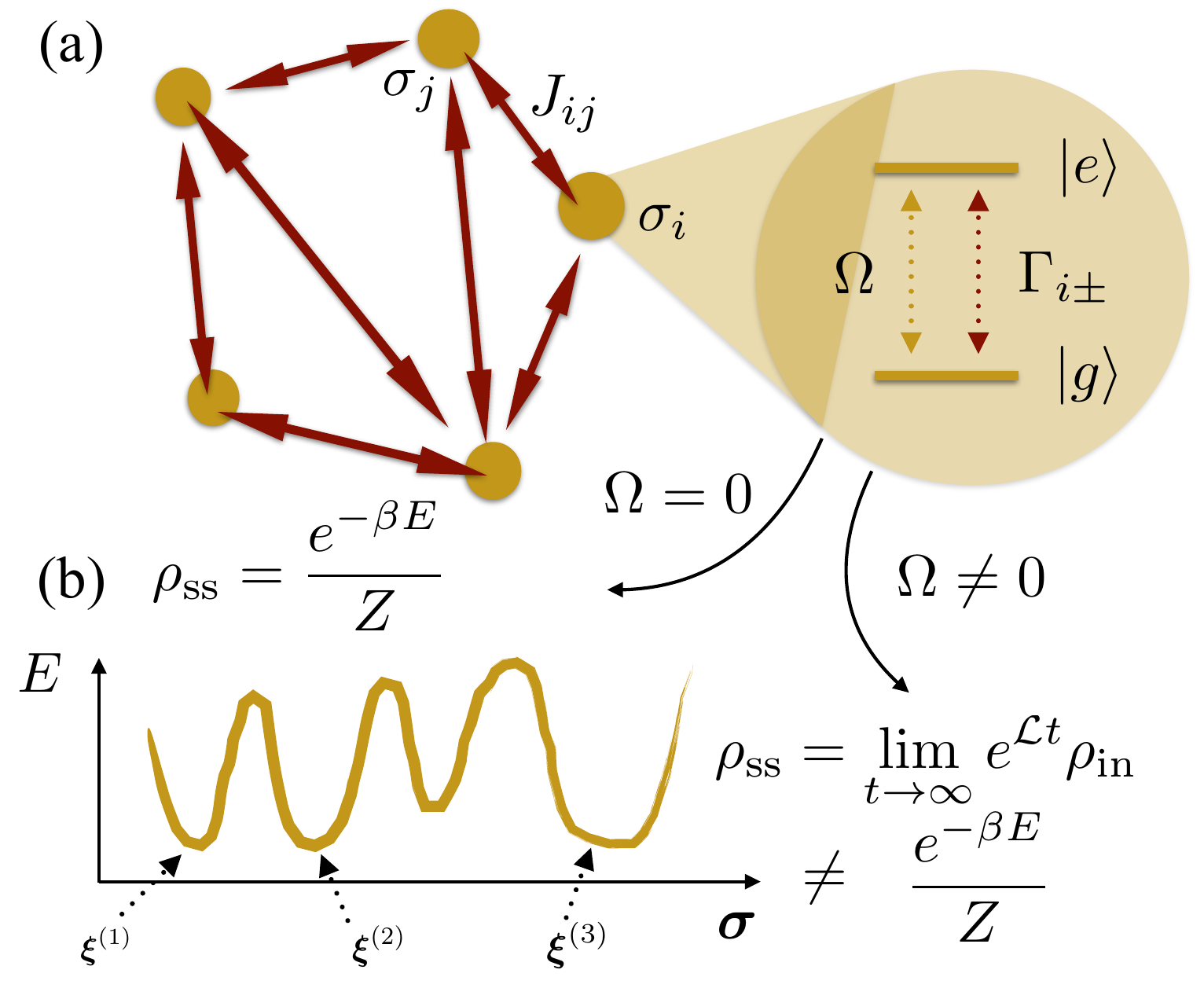}
\caption{\textbf{Sketch of the classical-to-quantum mapping for the Hopfield NN.} (a) In the Hopfield model neurons (dots) are binary spins describing the activity of the neurons (+1 firing, -1 silent). The OQSs framework allows us to study the competition between thermal and quantum effects. In particular, the $i$-th neuron changes its activity state at a rate $\Gamma_{i\pm}$ as in the classical model or undergoes a quantum state change, due to the coherent driving introduced in Eq.\ \eqref{driving}. (b) If $\Omega = 0$, the stationary state is at thermal equilibrium. The qualitative behavior of the  energy function of the classical NN is sketched in a one dimensional projection of the configurational space. Memory patterns are stored as the energy minima of the energy function. Whenever the NN is initialized close enough (close in the sense of the \emph{Hamming distance} between spin configurations) to a specific memory pattern, the dynamics in Eq.\ \eqref{dynHopf} allows to retrieve the corresponding stored pattern. In the presence of quantum effects ($\Omega \neq 0$), the nature of the stationary state can be non-trivial, i.e. it may be non-thermal, due to the competition between quantum coherence and irreversible classical dynamics.}
\label{Fig1}
\end{figure}

In order to make our ideas concrete we introduce a generalisation based on OQSs of one of the most studied NN systems, the Hopfield model \cite{Hopfield:PNAS:1982} (see Fig. \ref{Fig1}). The dissipative part of the dynamics corresponds to the thermal stochastic dynamics of the classical  model (often realised via classical Monte Carlo dynamics), while quantum effects are due to a transverse field Hamiltonian that turns this classical equilibrium system into a quantum non-equilibrium one. In this approach, the result of a NN computation is imprinted on the long time density matrix, whose properties as a function of the control parameters (temperature, quantum driving, initial state) determine the phase diagram of the system.  This setup reduces to the classical Hopfield NN when the quantum Hamiltonian is removed.  By means of mean-field methods (which are exact for fully connected models such as this one) we calculate the phase diagram of the model, and show that a new non-equilibrium phase, characterized by the presence of limit cycles (LCs), arises due to the competition between coherent and dissipative dynamics. This may be regarded as a quantum generalisation of the retrieval phase of the classical Hopfield model.

Originally, the Hopfield NN was introduced as a toy model of \emph{associative memory}. In the human brain memory patterns are supposed to be retrieved by association. In a NN, this translates in the following: when a pattern similar enough to one of those stored is presented to the NN, the system is able to retrieve the correct one via classical annealing. The two fundamental ingredients to reproduce this are: (i) a dynamics on a system of $N$ binary spins ($\sigma_i = \pm 1$, $i = 1,\dots,N$), that represent neuron activity ($+1$ firing and $-1$ silent); (ii) an appropriate prescription for the couplings $J_{ij}$ that connect the $i$-th neuron with the $j$-th one, which must be able to store a set of $p$ different memory patterns $\xi_i^{(\mu)}$ (i.e. fixed spin configurations) with $i=1,\dots,N$, $\mu = 1, \dots, p$ . In this language, \emph{memory retrieval} denotes a phase in which the dynamics drives the system towards configurations which are closely resembling one of the $\xi_i^{(\mu)}$ for some $\mu$. It turns out that the following discrete time asynchronous dynamics fulfills these requirements:
\begin{equation}
\label{dynHopf}
\sigma_i (t+1) = \text{sign} \left(\sum_{j\neq i}J_{ij} \sigma_j (t)\right)\,, \quad J_{ij} = \frac{1}{N} \sum_{\mu=1}^p \xi_i^{(\mu)} \xi_j^{(\mu)}\,.
\end{equation}
Indeed Eq.\ (\ref{dynHopf}) describes a zero temperature Monte Carlo dynamics (that can be easily generalized to include thermal effects \citep{Amit:PRA:1985}). Moreover it can be proven that this dynamics minimizes the energy function $E= -\frac{1}{2} \sum_{i \neq j} J_{ij} \sigma_i \sigma_j$, namely an Ising model with pattern-dependent couplings and the global minima of $E$ are precisely the memory patterns $\xi_i^{(\mu)}$ (as long as $p \ll N$). Techniques used in the statistical physics of disordered systems enable to investigate Hopfield NNs (and more general types of NNs) quantitatively \cite{Amit:PRA:1985, Amit:PRL:1985, Amit:AnnPhys:1987}. In statistical physics language, the retrieval phase is the low temperature phase corresponding to an energy landscape where memory patterns are stable states of the NN, i.e.\ the thermal equilibrium stationary states; see Fig. \ref{Fig1}.

\section{Methods}

\subsection{Open quantum Hopfield NNs} 

To introduce quantum effects, we employ a description of the NN dynamics in terms of open quantum master equations. The starting point of our analysis is a master equation in Lindblad form for the density matrix $\rho$:
\begin{equation}
\dot{\rho} = -i [H, \rho] + \sum_{i=1}^N\sum_{\tau=\pm}\left( L_{i \tau}^{\phantom{\dagger}} \rho L_{i\tau}^{\dagger} - \frac{1}{2} \{L_{i\tau}^{\dagger}L_{i\tau}^{\phantom{\dagger}},\rho\}\right)\,,
\label{lindblad}
\end{equation}
where we define a set of jump operators as follows:
\begin{equation}
L_{i \pm} = \Gamma_{i \pm} \sigma_i^{\pm}, \quad \Gamma_{i\pm} = \frac{e^{\mp \beta/2 \Delta E_i}}{(2 \cosh(\beta \Delta E_i))^{\frac{1}{2}}}\,.
\label{jump}
\end{equation}
Here $\beta = 1/T$ is the inverse temperature, $\Delta E_i = \sum_{j\neq i} J_{ij}\sigma_{j}^z$ the change in energy under flipping of the $i$-th spin, and $\sigma_i^{\pm} =(\sigma_i^{x} \pm i \sigma_i^{y})/2$, with $\sigma^{x,y,z}$ the Pauli matrices. Quantum effects are included by a uniform transverse field in the $x$-direction, corresponding to a Hamiltonian,
\begin{equation}
H = \Omega \sum_{i=1}^N \sigma_i^x\,.
\label{driving}
\end{equation}
In the absence of this term, Eq.\ \eqref{lindblad} describes a classical stochastic dynamics: any initial density matrix that is diagonal in the $\sigma^z$ basis remains diagonal under the evolution and Eq.\ \eqref{lindblad} reduces to
$\dot{P} = \sum_{i=1}^N\sum_{\tau=\pm}
\Gamma_{i \tau}^2 \left[ \sigma_i^\tau - \frac{1}{2} \left( 1 + \tau \sigma_i^z \right) \right] P$, where $P$ is the probability vector formed by the diagonal of $\rho$. The rates $\Gamma_{i \tau}^2$ obey detailed balance  with respect to the Boltzmann distribution for energy $E$ at temperature $T$, so that this is the master equation for the classical Hopfield NN, as shown in the supplementary material (SM) \cite{SM}.

\subsection{Derivation of the mean field equations} 

Starting from Eq. \eqref{lindblad}, the corresponding equation for the evolution of an observable $O$ is given by:
\be
\begin{split}
	\dot{O} & = i [H,O] + \sum_{k} L^{\dagger}_{k} O L_{k} - \frac{1}{2}\{L^{\dagger}_{k} L_{k},O\}\,,
\end{split}
\ee
where $H$ is given in Eq. \eqref{driving} and the jump operators are those in Eq. \eqref{jump}. Here we are interested in the evolution of the local observables $\sigma^{\alpha}_i$ ($\alpha=x,y,z$). The equation of motion for $\sigma^z_i$ reads (see SM for more details):
\be
	\dot{\sigma}^z_i = 2 \Omega \sigma^y_i - \gamma \sigma_i^z  + \gamma \tanh \lt \beta \Delta E_i \rt,
\ee
which in the absence of dissipation ($\gamma =0$) would simply describe Rabi oscillations of frequency $2\Omega$ about the $x$ axis. This clearly shows that the decoupling of the classical subspace does not hold any more: in fact, the equations for $\sigma_z^i$ do not close and we need to write additional ones for $\sigma_i^{x/y}$ or, equivalently, $\sigma_i^\pm $. These are generically more involved, since these operators do not commute with the $\Gamma$s. To simplify the respective Lindblad equations, we will make use in the following of the anticommutation relations, which in particular imply that
\begin{equation}
f(\sigma_1^z,\cdots,\sigma_j^z,\cdots, \sigma_N^z) \sigma^{\pm}_j = \sigma^{\pm}_j f(\sigma_1^z,\cdots,-\sigma_j^z,\cdots, \sigma_N^z)\,, 
\label{exc}
\end{equation}
for any function $f$ with a power series representation. Let us consider the equation for $\sigma_i^+$:
%
%In the quantum case a coherent dynamics is introduced through an Hamiltonian term, coupling the diagonal and non-diagonal terms of the density matrix, namely: $H = \Omega \sum_{i=1}^N \sigma_i^x$. The equation of motion for $\sigma^z_i$ is local (where local is intended as above):
%\begin{equation}
%\dot{\sigma}^z_i = 2 \Omega \sigma^y_i + \gamma (1-\sigma^z_i) \Gamma_{i+}^2 - \gamma(1+\sigma^z_i) \Gamma_{i-}^2\,.
%\end{equation}
%The equations for $\sigma_i^{\pm}$ are more involved, since these observables do not commute with the non-trivial operatorial part of the Lindblad jump operators, i.e. $[\Gamma_{i\sigma},\sigma_l^{\pm}] \neq 0$. In order to make some progress, the following property is used:
%\begin{equation}
%f(\sigma_1^z,\cdots,\sigma_j^z,\cdots, \sigma_N^z) \sigma^{\pm}_j = \sigma^{\pm}_j f(\sigma_1^z,\cdots,-\sigma_j^z,\cdots, \sigma_N^z)\,, 
%\label{exc}
%\end{equation}
%which is valid for any function $f$ with a power series representation. Let us consider the equation for $\sigma_i^+$:
\begin{align}
\dot{\sigma}_i^+ = -i\Omega \sigma_i^z +\gamma\sum_{l} \sigma_l^- \Gamma_{l+} \sigma_i^+ \Gamma_{l+}\sigma_l^+  
- \frac{\gamma}{2} \sum_l \left(\{\sigma_l^- \Gamma_{l+} \Gamma_{l+}\sigma_l^+,\sigma_i^+\} + (+ \leftrightarrow -)\right)\,.
\label{eqsigmap}
\end{align} 
Our aim is now to move all the $\Gamma$'s to the right. This is done applying property (\ref{exc}). Introducing the notation
\begin{equation}
\Gamma_{l s}^{(i)} = \Gamma_{ls}(\sigma_1^z,\cdots,-\sigma^z_{i},\cdots,\sigma^z_N)\,,
\end{equation}
where $s \in \set{+,-}$. Eq. (\ref{eqsigmap}) can be written in a compact form as
\begin{align}
\dot{\sigma}^+_i &= -i\Omega \sigma_i^z - \frac{\gamma}{2}\sigma_i^+ (\Gamma_{i+}^2+\Gamma_{i-}^2) 
+\frac{\gamma}{2}\sigma^+_i \sum_{l\neq i} (1-\sigma_l^z) \lt \Gamma_{l+}^{(i)} \Gamma_{l+} -\frac{1}{2}\Gamma_{l+}^{(i)}\Gamma_{l+}^{(i)}-\frac{1}{2}\Gamma_{l+}\Gamma_{l+} \rt+ \nol[2mm]
&+\frac{\gamma}{2}\sigma^+_i \sum_{l\neq i} (1+\sigma_l^z) \lt \Gamma_{l-}^{(i)} \Gamma_{l-} - \frac{1}{2}\Gamma_{l-}^{(i)}\Gamma_{l-}^{(i)} - \frac{1}{2}\Gamma_{l-}\Gamma_{l-} \rt \,. 
\label{eq:sigmap}
\end{align}
The equation for $\sigma_i^-$ is simply obtained via hermitian conjugation of the one above. These equations can be simplified recognizing that $\Gamma$ and $\Gamma^{(i)}$ depend on two configurations which differ by a single spin (the $i$-th one). It could be thereby reasonably expected that $\Gamma \approx \Gamma^{(i)}$ up to finite-size corrections which scale as $1/N$ (\textbf{this is true as long as $p \ll N^2$}, see also SM for more details).  Therefore, we can safely neglect these terms in the thermodynamic limit, which produces a considerably simplified form of the dynamical equations
\be
	\dot{\sigma}_i^\pm = \mp i\Omega \sigma^z_i - \frac{\gamma}{2} \sigma_i^\pm.
\ee
The equations can be further simplified by expressing them in the $(x,y,z)$ basis:
\begin{subequations}
\begin{align}
	\dot{\sigma}_i^z &= 2 \Omega \sigma_i^y - \gamma \sigma_i^z + \gamma \tanh\lt \beta \Delta E_i \rt, \\
	\dot{\sigma}_i^y &= -2 \Omega \sigma_i^z - \frac{\gamma}{2} \sigma_i^y, \\
	\dot{\sigma}_i^x &=  - \frac{\gamma}{2} \sigma_i^x .
\end{align}
\end{subequations}
Interestingly enough, the equation for $\sigma_i^x$ decouples from the others and one can conveniently restrict to the $y$ and $z$ components only.

At this stage, we define a suitable set of collective variables as:
\begin{equation}
s_{\mu}^{\alpha} = \frac{1}{N} \sum_{i=1}^N \xi_{i}^{(\mu)} \sigma_i^{\alpha} \quad \mu=1,\cdots,p\,; \,\,\,\,\, \alpha = x,y,z\,.
\end{equation} 
These observables can be thought as the overlap between the $\mu$-th memory and the $\alpha$ component of the spins. 
%More importantly, the resulting equations for these variables are closed, thus considerably reducing the computational difficulty of the problem from $2N$ to $2p$ coupled equations. 
The equations for these collective variables read
\begin{align}
\dot{\mathbf{s}}^z &= 2\Omega \mathbf{s}^y +  \frac{\gamma}{N} \sum_{i=1}^N \pmb{\xi}_i \tanh(\beta \pmb{\xi}_i \cdot \mathbf{s}^z ) -\gamma \mathbf{s}^z\,,\\
\dot{\mathbf{s}}^y &= -2\Omega \mathbf{s}^z -\frac{\gamma}{2} \mathbf{s}^y\,,
\end{align} 
where the vectorial notation now groups the pattern indices, not the positional ones: in other words, these are $p$-uples of operators $\mathbf{s}^\alpha = \lt s^\alpha_1 , \ldots s^\alpha_p   \rt^\intercal$ and numbers $\pmb{\xi}_i = \lt \xi^{(1)}_i , \ldots \xi^{(p)}_i   \rt^\intercal$.

%Eqs. (\ref{lindblad}-\ref{driving}) define our open quantum Hopfield model.

\section{Results}

\subsection{Mean-field solution} 

Since the NN is defined in terms of fully-connected interactions, the mean-field approximation should be exact \cite{SM} and amounts to considering the evolution of the averaged collective variables $m_{\mu}^{z,y}=\braket{s_{\mu}^{z,y}}$ neglecting correlations between them \cite{SM}. The evolution equations then read:
\begin{align}
\dot{\mathbf{m}}^z &= 2\Omega \mathbf{m}^y +  \frac{1}{N} \sum_{i=1}^N \pmb{\xi}_i \tanh(\beta \pmb{\xi}_i \cdot \mathbf{m}^z ) - \mathbf{m}^z\,,
\label{mfdyn1}
\\
\dot{\mathbf{m}}^y &= -2\Omega \mathbf{m}^z -\frac{1}{2} \mathbf{m}^y\,,
\label{mfdyn2}
\end{align}
where we introduced a vectorial notation for both the collective variables and the memory patterns, $\dot{\mathbf{m}}^\alpha = \left( m_1^\alpha , \ldots , m_p^\alpha \right)$. $m^z_{\mu}$ represents the overlap between the $\mu$-th memory pattern and the neuronal configuration of the system (in the $z$ direction), and thus can be used as an \emph{order parameter}: if in the stationary state $m_1^z \approx (1,0,\dots,0)$, this means that the NN correctly retrieved the first pattern stored (and similarly for the other $p-1$ patterns). Setting $\dot{\mathbf{m}}^z=\dot{\mathbf{m}}^y= \mathbf{0}$ in Eqs.\ (\ref{mfdyn1},\ref{mfdyn2}) we get the equations for the stationary solutions:  
\begin{equation}
(1+8\Omega^2)\mathbf{m}^z = \overline{
\pmb{\xi} \tanh (\beta \pmb{\xi} \cdot \mathbf{m}^z)
}
\,,
\label{stationary}
\end{equation}
where we have assumed \emph{self-averaging}, typical of disordered systems \cite{Mezard:book} in the large $N$ limit, 
i.e., $1/N \sum_{i=1}^N f(\pmb{\xi}_i)\rightarrow \overline{f(\pmb{\xi})}$, where $\overline{(\cdot)}$ is the average over the disorder distribution. Equations (\ref{mfdyn1},\ref{mfdyn2}) allow to study the interplay between the retrieval and the paramagnetic phases of the NN. It is worth remarking, however, that a more general approach, involving the replica method, is required to deal with the spin glass phase arising when an extensive number of memory patterns is loaded into the NN \cite{Amit:PRL:1985}. In the following we focus on the retrieval phase of our OQS, which amounts to considering the case of a finite number of patterns $p$.

As a first step, we notice that Eq.\ \eqref{stationary} has the same form of the mean-field equation obtained for the classical Hopfield NN  \cite{Amit:PRA:1985}: in particular a suitable rescaling of $\mathbf{m}^z$ with an effective temperature $T_{\rm eff} = T (1+8\Omega^2)$ establishes their equivalence. This means that the structure of the stationary points of the dynamics is equivalent to the classical one up to a rescaling of the temperature. In particular, the retrieval solutions $\mathbf{m}^z \approx (1,\dots,0)$, the paramagnetic solution $\mathbf{m}^z =\mathbf{0}$, as well as the whole set of metastable states (spurious memories) \cite{Amit:PRA:1985} are still fixed points of the quantum dynamics.
Closer inspection of Eqs.\ (\ref{mfdyn1},\ref{mfdyn2}) reveals, however, that quantum driving does more than just rescale temperature.

\begin{figure}[t]
\includegraphics[width=0.48 \textwidth]{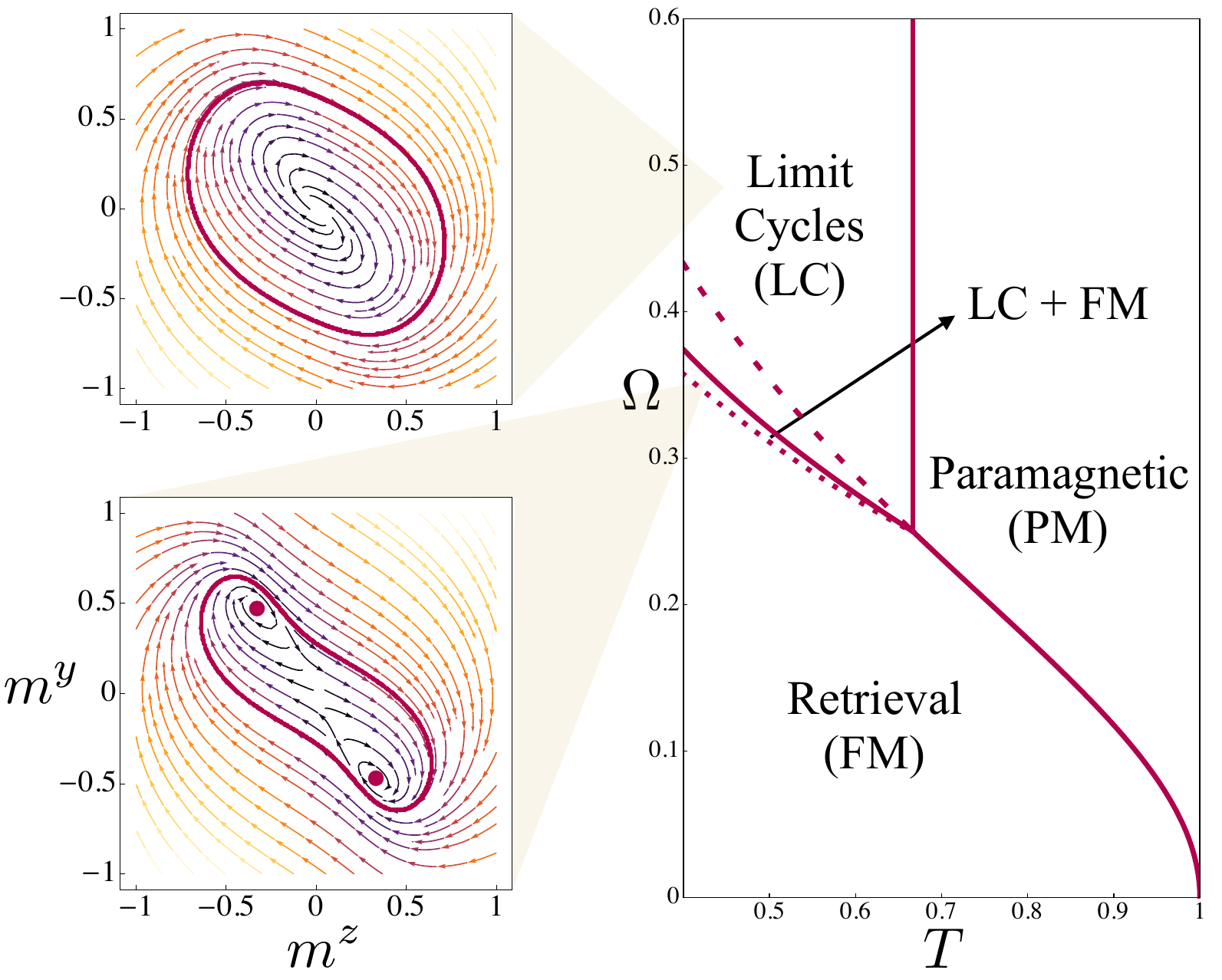}
\caption{\textbf{Phase diagram of the OQS generalization of the Hopfield model in the $(T,\Omega)$ plane.} The NN displays a paramagnetic phase in the high temperature regime, with a stable attractor in $\mathbf{m}^z=\mathbf{0}$. The boundary of the paramagnetic phase is established by looking at the stability of this attractor. In the low $T$ and $\Omega$ regime the system is in the ferromagnetic (retrieval) phase, whereas for low $T$ but sufficiently large $\Omega$, the NN displays a LC (top left), which is the only stable attractor once the ferromagnetic solutions become unstable (above the solid line between FM and LC phases). However the LC appears well below this line (and above the dotted line), producing a small region of phase coexistence (see flux diagram at bottom left). Above the dashed line and for $T <2/3$, the ferromagnetic solutions disappear.}
\label{Fig2}
\end{figure}

Beyond these stationary solutions, the mean-field equations (\ref{mfdyn1},\ref{mfdyn2}) also display time-dependent periodic solutions at long times.  A simple way to gain insight on this new feature is to consider a high-$T$ expansion of Eqs.\ (\ref{mfdyn1},\ref{mfdyn2}) up to the first non-linear order.  In this case the equations become analogous to a Lotka-Volterra dynamical system, widely studied in the literature on ecological systems \cite{Takeuchi:book}. Under appropriate conditions, a Lotka-Volterra system is known to have limit cycles (LCs) solutions.

The observation above suggests that our open quantum Hopfield NN can feature at least three possible phases in the $(T,\Omega)$ plane: (i) a paramagnetic phase where the dynamics converges to the trivial solution $\mathbf{m}^z = \mathbf{0}$; (ii) a retrieval phase where the attractor of the system is one of the non-trivial solutions of Eq.\ \eqref{stationary}; (iii) a novel time-dependent stationary phase, emerging from the competition between the dissipative and coherent dynamics.

\subsection{Phase diagram} 

To characterize the phase diagram of this OQS in the $(T,\Omega)$ plane, we follow two different routes: (i) we solve numerically the mean-field equations at low $p$ by averaging over disorder using the factorized probability distribution $P(\pmb{\xi}) = \prod_{\alpha=1}^p p (\xi^{(\alpha)})$ with $p(\xi) = 1/2\delta (\xi-1) + 1/2\delta (\xi+1)$; (ii) we perform a \emph{Lyapunov linear stability analysis} \cite{Lyapunov:IJC:1992}, namely we study the dynamical stability of the stationary points under small perturbations (see \cite{SM} for details). Through either approach we identify the boundaries of the different phases. The numerical solution of the mean field equations \eqref{mfdyn1}, \eqref{mfdyn2} is obtained with a standard numerical integrator available in Wolfram Mathematica.

%The second method, in particular, amounts to linearising the system around the stationary state solutions and to evaluate the sign of the corresponding eigenvalues. A straightforward analysis shows that there are only two different eigenvalues $\lambda_{\pm}$ with degeneracy $p$ (see also SM):
%\begin{equation}
%\lambda_{\pm} = \frac{\kappa (\beta,\mathbf{m}^{\ast})-3/2 \pm \sqrt{(\kappa (\beta, \mathbf{m}^{\ast})-1/2)^2-16\Omega^2}}{2}\,,
%\label{eigenv}
%\end{equation}
%where $\kappa (\beta, \mathbf{m}^{\ast}) = \beta (1-\tanh^2{(\beta \mathbf{m}^{\ast})})$ and $\mathbf{m}^{\ast}$ indicates the generic stationary state solution.
The phase diagram of the NN is summarized in Fig. \ref{Fig2}. For $T>2/3$ and $\Omega > \sqrt{(1-T)/8T}$ the system is in the paramagnetic phase. Retrieval is possible in the low temperature and low $\Omega$ regime, whereas the dynamics features LCs as a stationary manifold in the low temperature and high coherence regime. At the boundary between the retrieval and the LC phase, we recognize a small region where the two phases coexist: initializing the system close enough to $(\mathbf{m}^z,\mathbf{m}^y) = (\mathbf{0},\mathbf{0})$, drives the NN towards the retrieval solutions, whereas initial conditions chosen outside a critical hypervolume centered around $(\mathbf{0},\mathbf{0})$ converge to the LC (see insets in Fig \ref{Fig2}).

\begin{figure}[t]
\includegraphics[height=7cm, width=8.7cm]{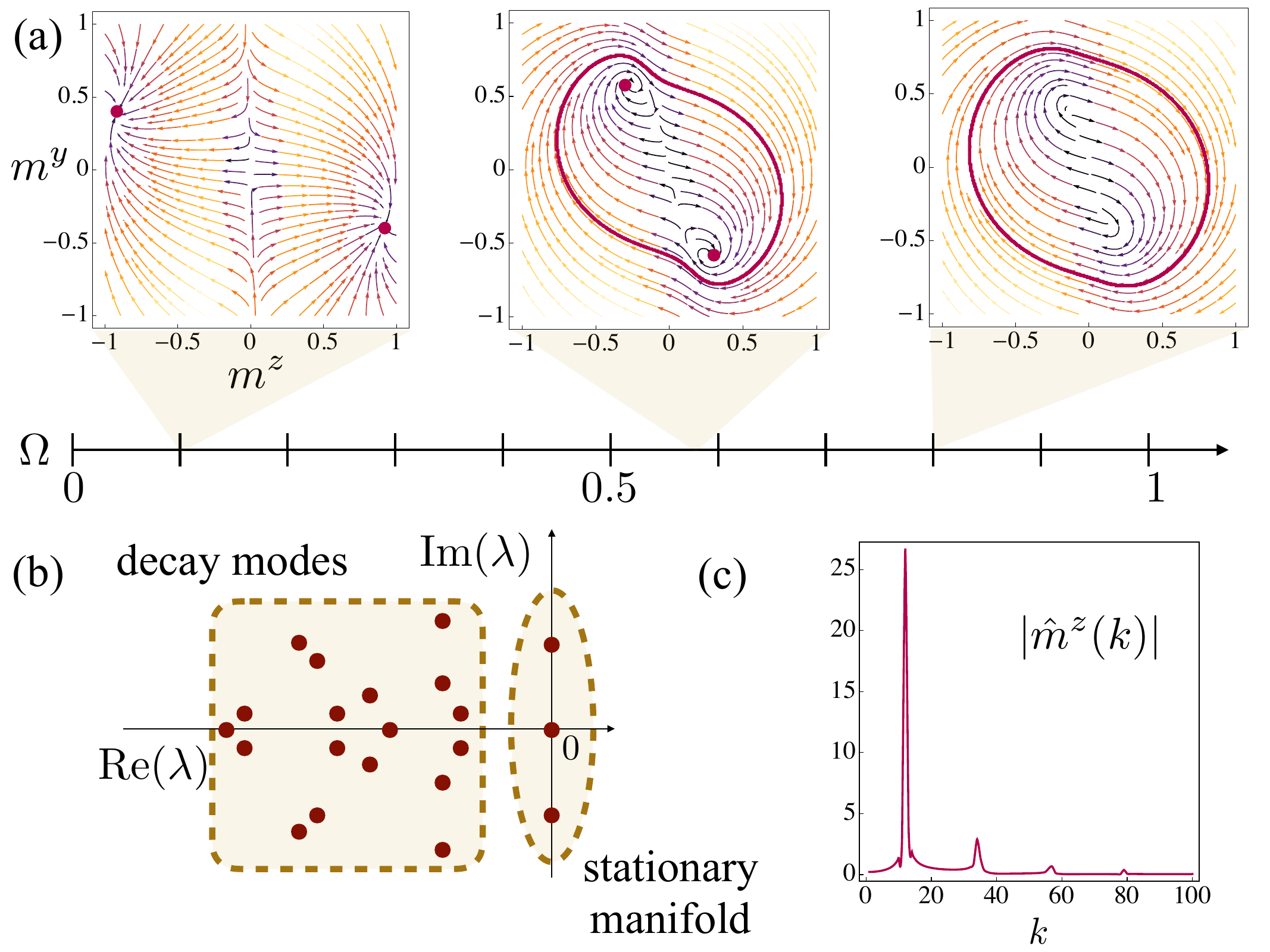}
\caption{\textbf{Qualitative spectrum of the superoperator in the LC phase.} (a) From left to right flux diagrams of the mean-field dynamics at $T=0.15$, $\Omega = 0.1\,,0.58\,,0.8$. The different attractors are highlighted in each panel. In the bulk of the LC phase the maximum overlap with one of the stored patterns is $\sim 0.8$, independently from the initial conditions, suggesting that also this phase can be interpreted as a retrieval phase. (b) Sketch of the spectrum of a plausible Liovillian super-operator describing the LC shown in (a). Time-dependent periodic stationary states require at least a conjugate pair of purely imaginary eigenvalues. (c) The structure of the Fourier components $|\hat{m}^z(k)|$ of the top right LC, justifies the single frequency approximation made in the main text.}
\label{Fig3}
\end{figure}

For a single pattern, $p = 1$, we are able to rigorously prove the presence of a LC phase, since Eqs.~\eqref{mfdyn1} and \eqref{mfdyn2} can be recast in \emph{Li\' enard form} \cite{Palit2010}, (see also \cite{SM}) for $T<2/3$ and $T> (8\Omega^2 + 1)^{-1}$. The Li\' enard theorem \cite{Hirsch_book} guarantees the existence, uniqueness and stability of a LC (which wraps around the origin).
%Outside this region, our numerical analysis of the mean-field equations shows the persistence of the limit cycle up to a threshold exceeding the stability line of the retrieval solutions, implying the presence of a regime of coexistence of stable retrieval points and a stable limit cycle.

The LC phase can be interpreted as a new quantum retrieval phase: for low enough temperatures, in fact, independently from the initial conditions the NN is always driven towards a LC with a large oscillation amplitude. For instance, the case reported in Fig.~\ref{Fig3}a (corresponding to $T = 0.15$) shows oscillations reaching a maximum overlap of $\sim 0.8$ with the single stored pattern. Limit cycles of this kind (involving mostly a single overlap) also exist for generic $p>1$. For $p \leq 4$ we have performed a systematic numerical study in the LC phase in the range of parameters $0.4 \leq T \leq 0.66$ and $0.4 \leq \Omega \leq 1$ (using the same distribution of the couplings mentioned above) finding that at long times oscillations involve at most two overlaps, while all the remaining $p-2$ asymptotically vanish. More importantly, starting from an initial condition with large overlap with one of the patterns (say, the $\mu$-th one) and significantly smaller overlap with the remaining ones seems to always lead to oscillations in the $(m^z_{\mu},m^y_{\mu})$ plane and vanishing ones in the others. These findings are also supported by preliminary investigations for $p < 10$ and seem plausible for general $p$ by looking at the theoretical argument presented in section 3C of the SM. This is somewhat reminiscent of the retrieval phase, where the dynamics proceeds towards the pattern it initially had larger overlap with. Furthermore, when cooling the system at large $\Omega$, LC attractors emerge before the appearance of the stable fixed points, which hints at the possibility of a correspondence between fixed point attractors $\xi_i^{(\mu)}$ and LC ones. At this level, this aspect comes out as a numerical finding and will require future work in order to be explored and established in more detail.
%
%
%
%
%This hints at the possibility of associating to any stable fixed point attractor $\xi_i^{(\mu)}$ of the NN a corresponding LC attractor, \textbf{but definitely asks for a deeper investigation that we will perform elsewhere}. 
It is moreover noteworthy that the fundamental properties of the Hopfield dynamics are rather robust against the introduction of a coherent term such as the transverse field we use in this work. It is worth remarking that this robustness is closely related to symmetry arguments: indeed the generalized $\mathrm{Z}_2$ symmetry broken in the FM phase of the classical model is still broken in the LC phase. Here, additionally, the stationary state spontaneously breaks the time translation symmetry, which should correspond to the appearance of a Goldstone mode, as found in \cite{Chan:PRA:2015}.

% whose amplitude can reach overlap up to $\sim 0.8$ with the single stored pattern. We conjecture that such phase also exists for general p. Numerical solutions of Eq.\ \eqref{mfdyn} appear to confirm this and show that upon initializing the system in a configuration which has zero overlap with all the memories but the $k$-th, the NN reaches a stationary state where the LC oscillates in the $2$-dimensional plane $(m^z_{\mu},m^y_{\mu})$, and also in this case the amplitude of the LC can grow up to $\sim 0.8$ (see Fig. \ref{Fig3}a). This means that it is possible to associate to any stable fixed point attractor $\xi_i^{(\mu)}$ of the NN a corresponding LC attractor, which can be thought as an exotic memory pattern, since it is possible to read out up to the $80\%$ of the memory pattern from it.

\section{Discussion}

Examples of LC phases in classical Hopfield NNs with asymmetric couplings exist \cite{Coolen:PRA:1988, Evans:JPA:1989, Inoue:JPC:2011}. However, the LC phase that we find here is intimately related to the competition between coherent and dissipative dynamics. The persistence of oscillations in the long time limit implies the survival -- to a degree -- of quantum coherence. To substantiate this claim, we argue in the following what the structure of the {\em stationary manifold} of the Lindblad equation should be, following the  classification of Refs.\ \cite{Baumgartner2008,Albert:PRA:2014,Albert2016} for the case of systems with finite-dimensional Hilbert spaces. 

%In order to observe such a regime, the stationary density matrix has to preserve, at least in part, quantum coherence. This claim can be justified by inspecting the possible stationary state structures of time-independent Lindblad master equations, for which an exhaustive classification has been recently put forward \cite{Albert:PRA:2014}.

By formally integrating Eq.~\eqref{lindblad} we get $\rho(t) = e^{t \mathcal{L}} \rho_{\mathrm{in}}$, where $\mathcal L$ is the generator of the open quantum dynamics [i.e., $\mathcal{L}\rho$ is shorthand for the r.h.s.~of Eq.~\eqref{lindblad}], and $\rho_{\mathrm{in}}$ is the initial state. Assuming $\mathcal{L}$ to be diagonalizable,
%(i.e., non non-trivial Jordan blocks)
the density matrix can be further expanded as \cite{Macieszczak2016}:
\begin{equation}
\rho(t) =  \sum_{l=1}^{2^{2N}} c_l e^{t \lambda_l} R_l\,,
\end{equation}
where $\lambda_l$ are the eigenvalues, $R_l$ are the right eigenmatrices of $\mathcal L$, i.e. $\mathcal L R_l = \lambda_l R_l$, and $c_l$ the components of the initial state on these eigenmodes. Because of preservation of probability and positivity, the eigenvalues of $\mathcal L$ have non-positive real parts and are either real or come in complex conjugate pairs. The stationary manifold is constructed from all the $R_l$ for which $\mathrm{Re}(\lambda_l) = 0$; we define $n$ its dimension and reorder the eigenvalues such that the zero ones appear first. If $\lambda_l = 0 \ \forall \ l = 1,\ldots, n$ then each initial state maps to a state in the stationary manifold asymptotically and no time dependence survives at long times.  To allow for limit cycles, $\mathcal{L}$ must display at least a pair of conjugate, purely-imaginary eigenvalues $\pm i \omega$, as sketched in Fig.~\ref{Fig3}b. The long time evolution due to the presence of these eigenvalues is unitary and one could formally define a corresponding reduced Hamiltonian acting coherently on the stationary subspace \cite{Zanardi2014c,Albert2016,Macieszczak2016}.

A Fourier analysis deep in the LC phase (see Fig \ref{Fig3}c) justifies a single frequency approximation for the stationary state, leading to $\rho_{\mathrm{ss}} (t) \sim R_0 + R_1 + R_2 e^{i \omega t} + R_2^{\dagger} e^{-i\omega t}$, that allows to obtain a quantitative agreement with the numerics by requiring that the $R_i$'s are such that:  $\Tr \left(s^\alpha R_{0/1}\right) = 0$, $\Tr \left( s^z R_{2}\right) =\bar{m}^z $ and $\Tr \left( s^y R_{2}\right) = \bar{m}^y e^{i\epsilon} $. This choice reproduces a LC in the $(m^z, m^y)$ plane parameterized as $m^z (t) = \Tr \left( s^z \rho_{\mathrm{ss}} (t)\right) \sim \bar{m}^z \cos( \omega t)$ and $m^y (t) = \Tr \left( s^y \rho_{\mathrm{ss}} (t)\right) \sim \bar{m}^y \cos(\omega t + \epsilon)$ (with $\epsilon$ a proper real phase).

\section{Conclusions}

We proposed a framework based on open quantum systems to investigate quantum effects in the dynamics of neural networks. It is worth remarking that our implementation is conceptually different from recent results where Hopfield-like unitary evoultions have been found in the context of multimodal cavity quantum electrodynamics with disorder \cite{Goldbart:PRL:2011,Strack:PRL:2011,Rotondo:PRL:2015}. We applied this approach to an open system quantum generalization of the simplest model of associative memory, the Hopfield model. As in the classical case it is possible to use a mean-field treatment to determine the phase diagram. We identified a retrieval phase with fixed points associated to classical patterns and quantum effects can be accounted for by an effective temperature. We moreover found a novel phase characterized by limit cycles which are a consequence of the quantum driving. Our approach is a natural extension of the NN paradigm into the domain of open quantum systems. It shows that the resulting phase structure of such systems can indeed be richer than that of their classical counterparts. Future investigations are needed in order to clarify whether practical applications of NNs can benefit from quantum effects. For instance, a relevant question is whether the LC phase discovered here has some interpretation from the statistical learning theory viewpoint.

%The framework for open quantum NNs we introduce here is similar in spirit to that of recent works where environment engineering is exploited to recover the result of a quantum computation from the dark state of a dissipative dynamics \cite{Diehl:NatPhys:2009, Verstraete:NatPhys:2009, Caspar:PRA:2016, Schindler:NatPhys:2013}.

\vspace{0.3cm}
\section{Acknowledgments} 

The research leading to these results has received funding from the
European Research Council under the European Union's Seventh Framework Programme (FP/2007-2013) / ERC Grant Agreement No. 335266 (ESCQUMA), the H2020-FETPROACT-2014 Grant No.640378 (RYSQ), and EPSRC Grant No. EP/M014266/1. P.R. acknowledges funding by the European Union through the H2020 - MCIF No. 766442.

\bibliographystyle{apsrev4-1}

\bibliography{opQnet_arXiv}

\end{document}